\documentclass[aps,epsfig,rotate,showpacs]{revtex4}  
\usepackage{epsfig}
\usepackage{graphicx}
\usepackage{amsmath}

\begin{document} 

\title{Coherent Manipulation of Quantum Delta-kicked Dynamics: Faster-than-classical Anomalous Diffusion}
\author{Jiangbin Gong, Hans Jakob Woerner, and Paul Brumer}
\affiliation{Chemical Physics Theory Group, Department of Chemistry,
University of Toronto, 
Toronto, Canada M5S 3H6}
\date{\today}

\begin{abstract}Large transporting
regular islands are found
in the classical phase space of a modified kicked rotor system in which
the kicking potential is reversed after every two kicks.
The corresponding quantum system, for a variety of system parameters
and
over long time scales,
is shown to display energy absorption that is
significantly faster than that associated with the underlying
classical anomalous diffusion.
The results are of interest to both areas of quantum chaos and quantum
control.\end{abstract}

\pacs{05.45.Mt, 32.80.Qk, 05.60.-k}

\maketitle

The kicked-rotor system (KR) has long served as a paradigm 
for classical and quantum chaos
\cite{casatibook}.  
Its atom optics 
realization 
\cite{raizenetc}
makes it possible to directly compare
fundamental theoretical work with corresponding experimental studies. 
The KR 
is also of interest to a variety of other fields such as
molecular physics \cite{fishman,averbukh}, condensed matter
physics
\cite{casatiprl} and
quantum information \cite{facchi,georgeot}.

It is well known that quantum coherence suppresses classical chaotic (normal) diffusion.
The suppression mechanism is well understood in terms of
``dynamical localization" \cite{casatibook} in KR.
By contrast,
much less is known about quantum interference effects
in the case of classical anomalous diffusion,  a phenomenon induced by
the fractal boundary between regular and 
chaotic regions in phase space.  Of particular 
interest is
the quantum dynamics of delta-kicked systems where the
underlying classical chaos coexists with transporting regular islands, e.g.,
the accelerator modes in KR \cite{grigolini,raizen,zaslavsky99,zaslavsky02,schanz}.
In this case it is found that anomalous diffusion induced
by transporting islands causes the early breakdown of
quantum-classical correspondence (QCC) \cite{grigolini}
and enhances
fluctuations of dynamical localization length \cite{raizen,zaslavsky99,zaslavsky02}, 
and that quantum eigenstates in the semiclassical limit 
may  ignore classical phase space structures \cite{schanz}.

Motivated by our recent studies on coherent manipulation of
classically chaotic dynamics \cite{gong01,gong02},
we consider here a modified kicked-rotor model (MKR) in which
the kicking potential is reversed after every two kicks \cite{danapre}.
As shown below, this apparently slight modification of the KR system has a 
profound effect on the dynamics. First, it results in the appearance of 
transporting regular islands that are much larger and
of a  different nature than those previously observed.
Second, and more importantly, we find a new phenomenon: that
the corresponding quantum system, for a variety of system parameters
and over considerably long time scales,
displays energy absorption that is significantly {\it faster} than that associated
with the underlying classical anomalous diffusion.
This result constitutes  an important and intriguing aspect of quantum interference effects
in classically chaotic systems.  Furthermore,
the drastic difference
in quantum dynamics between  
KR and MKR is a demonstration of spectacular quantum control \cite{controlbooks}
over delta-kicked systems, achieved here by 
periodically reversing or delaying the kicking field.
These results are of broad theoretical and experimental interest.

The Hamiltonians for the KR and MKR systems are given by
\begin{eqnarray}
H^{KR}(\hat{L},\theta, t)=\hat{L}^{2}/2I+\lambda \cos(\theta)\sum_{n}\delta(\frac{t}{T}-n),
\end{eqnarray}
and 
\begin{eqnarray}
H^{MKR}(\hat{L},\theta, t)=\hat{L}^{2}/2I+\lambda \cos(\theta)\sum_{n}f(n)\delta(\frac{t}{T}-n).
\end{eqnarray}
Here $|f(n)|=1$ and it changes sign after every two kicks (i.e., $f(n)=1$ if
$n=4j+1$, or $4j+2$, and $f(n)=-1$ if $n=4j+3$ or $4j+4$, where $j$ is
an integer),
$\hat{L}$ is the angular momentum operator, $\theta$ is
the conjugate angle,  $I$ is the moment of inertia, $\lambda$
is the strength of the kicking field, and $T$ is the time interval between kicks.
The basis states of their Hilbert spaces are given by $|m\rangle$, 
with $\hat{L}|m\rangle=m\hbar|m\rangle$.  
The quantum propagator in the KR case 
for times $(N-0^{+})T$ to $(N+1-0^{+})T$ is given
by
\begin{eqnarray}
\hat{F}^{KR}(\tau,k)=\exp(i\frac{\tau}{2}\frac{\partial^{2}}{\partial \theta^{2}})
\exp[-ik\cos(\theta)],
\end{eqnarray}
with dimensionless parameters
$ k=\lambda T/\hbar$ and the effective Planck constant
$\tau=\hbar T/I$.
The classical limit of the KR quantum map, i.e., the standard map, depends only
on one parameter $\kappa\equiv k\tau$ and takes the following form:
\begin{eqnarray}
\tilde{L}_{N}&=&\tilde{L}_{N-1}+\kappa \sin(\theta_{N-1}),\nonumber \\ 
\theta_{N}&=&\theta_{N-1}+ \tilde{L}_{N}, \end{eqnarray} where
$\tilde{L}\equiv L\tau/\hbar$ is the scaled c-number angular momentum
and $(\tilde{L}_{N}, \theta_{N})$ represents
the phase space location of a classical
trajectory at $(N+1-0^{+})T$.
With similar notation, the quantum map associated with MKR, for times
 $(4j+1-0^{+})T$ to $[4(j+1)+1-0^{+}]T$,
can be written as 
\begin{eqnarray}
\hat{F}^{MKR}(\tau,k)=[\hat{F}^{KR}(\tau,-k)]^{2}[\hat{F}^{KR}(\tau,k)]^{2}.
\label{mkr-map}
\end{eqnarray}
The classical limit of the MKR quantum map [Eq. (\ref{mkr-map})] is given  by
\begin{eqnarray}
\tilde{L}_{4j+1}&=&\tilde{L}_{4j}+\kappa \sin(\theta_{4j}), 
\ \theta_{4j+1}=\theta_{4j}+ \tilde{L}_{4j+1}, \nonumber \\ 
\tilde{L}_{4j+2}&=&\tilde{L}_{4j+1}+\kappa \sin(\theta_{4j+1}), 
\ \theta_{4j+2}=\theta_{4j+1}+ \tilde{L}_{4j+2}, \nonumber \\ 
\tilde{L}_{4j+3}&=&\tilde{L}_{4j+2}-\kappa \sin(\theta_{4j+2}), 
\ \theta_{4j+3}=\theta_{4j+2}+ \tilde{L}_{4j+3}, \nonumber\\
\tilde{L}_{4j+4}&=&\tilde{L}_{4j+3}-\kappa \sin(\theta_{4j+3}), 
\ \theta_{4j+4}=\theta_{4j+3}+ \tilde{L}_{4j+4}. \label{cla-mkr4}
\end{eqnarray}

The well-known accelerator modes in KR 
are closely related to the existence of
the marginally stable points: $
(\tilde{L}=2\pi l_{1}, \theta=\pm \pi/2)$
for $\kappa=2\pi l_{2}$, where $l_{1}$ and $l_{2}$ are integers.  These points
are shifted by a constant value ($\pm 2\pi l_{2}$) in $\tilde{L}$ after each iteration.  
For values of $\kappa$ close to $2\pi l_{2}$, there still exist
interesting transporting regular islands on which classical trajectories
simply jump to other similar islands located in adjacent phase space cells, resulting in a ballistic
increase of rotational energy.  For trajectories initially outside
the accelerator modes, the ``stickiness'' of the boundary between
the accelerator modes and the chaotic sea induces
anomalous diffusion over the energy space, i.e., energy increases
in a nonlinear fashion, but not quadratically.

\begin{figure}[ht]
\begin{center}
\epsfig{file=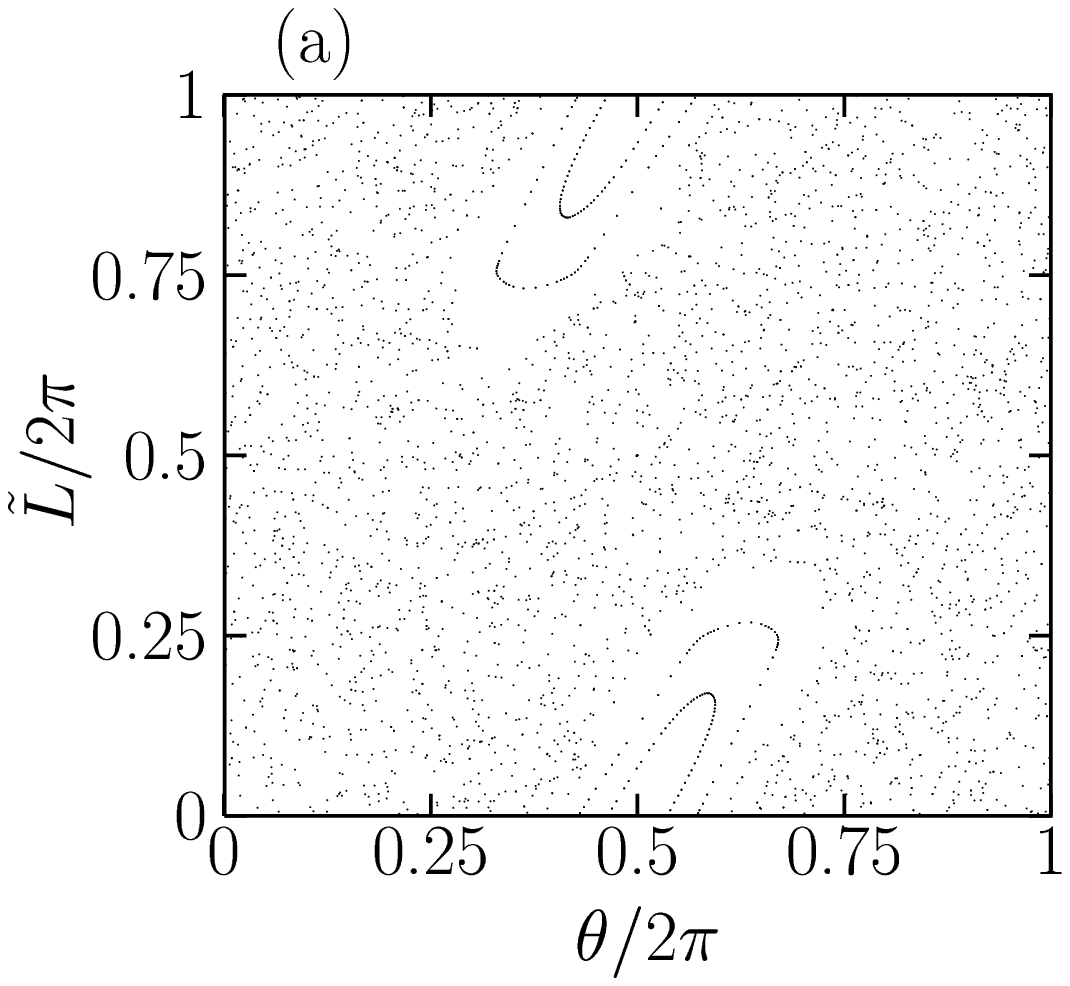,width=7.2cm}
\hspace{-0.3cm} \epsfig{file=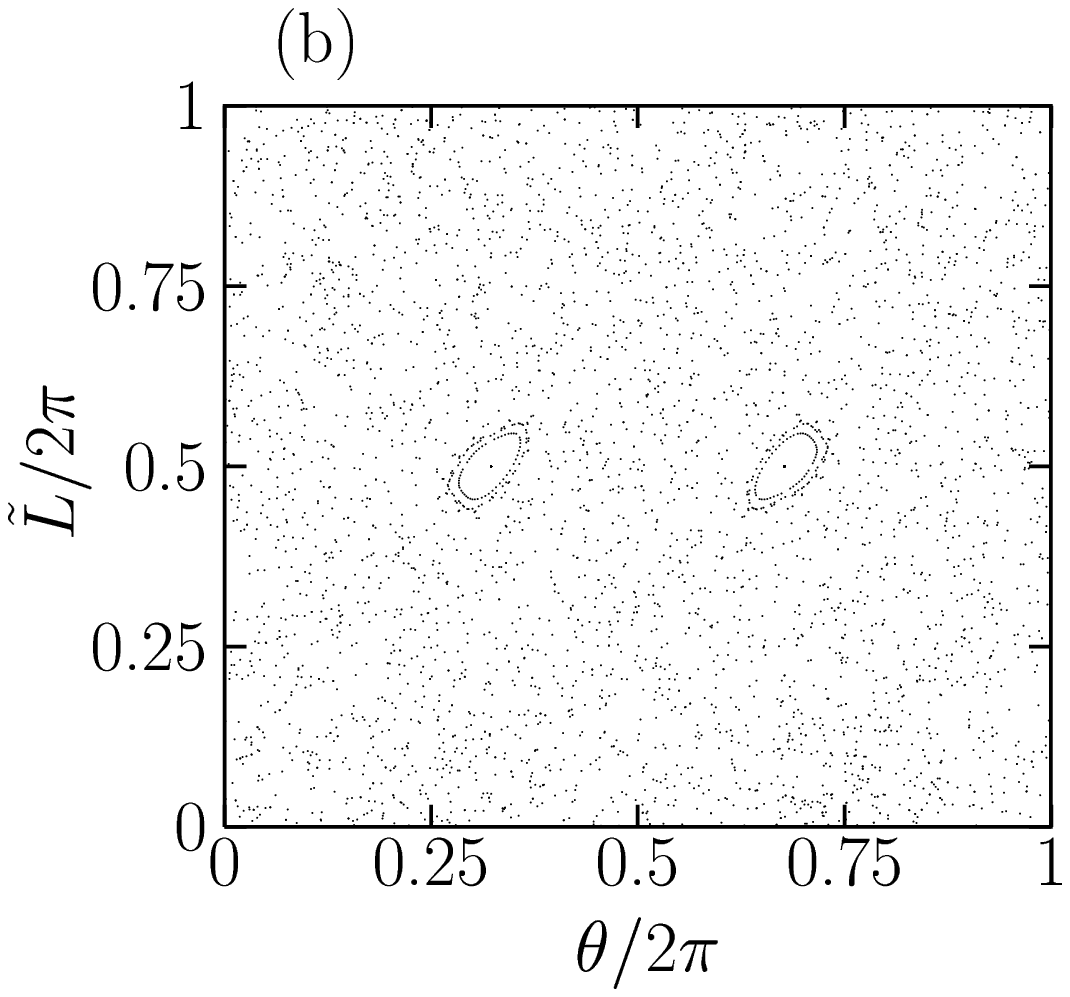,width=7.2cm}
\end{center}
\caption{Classical phase space structures for (a) the standard map and
(b) the map of Eq. (\ref{cla-mkr4}), in the case of $\kappa=3.5$. All variables are
in dimensionless units.
Note that the regular islands seen in (b) are
transporting while those in (a) are not.}
\label{clamap}
\end{figure}

Examining the classical MKR map 
[Eq. (\ref{cla-mkr4})],
we find 
a different type of marginally stable points. That is, 
for $\kappa=(2l_{2}+1)\pi$, trajectories emanating from
$\tilde{L}=(2l_{1}+1)\pi, \theta=\pm \pi/2$ 
are shifted by a constant value ($\pm (2l_{2}+1)\pi$) in $\tilde{L}$ 
after each kick.  As confirmed by our numerical experiments, this suggests 
that by reversing the kicking potential
after every two kicks  
one may create new transporting regular islands.
For example,
in Fig. \ref{clamap} we show classical phase space structures of both KR  
and MKR for $\kappa=3.5$.  The main regular
islands seen in Fig. \ref{clamap}a in the KR case are not transporting, consistent  with
the fact that $\kappa=3.5$ is far from 
$2\pi l_{2}$.  By contrast, a simple computation reveals that
those islands clearly seen in  Fig. \ref{clamap}b
in the MKR case 
are transporting (note that $\kappa=3.5$ is not far from $\pi$).  
A number of additional remarks are in order:  
(1) 
The transporting islands shown in Fig. \ref{clamap}b are much larger than
the well-studied accelerator modes in KR. 
A rough estimate gives their area  
at least 10 times larger than the accelerator modes in KR with
$\kappa=6.908745$, and 4 times larger with $\kappa=6.476939$ \cite{zaslavsky02};
(2) The transporting islands
are stable with changes in $\kappa$, e.g., the $\kappa=3.4$ or $\kappa=3.6$ case gives
transporting islands of similar size;
(3) Unlike the accelerator modes in KR, 
the transporting islands of MKR are far away from $\tilde{L}=0$. 
This latter feature favors the observation of
anomalous diffusion,  insofar as 
any initial state with sufficiently low energy will not
overlap with the transporting islands of MKR;
(4) After each kick,
trajectories initiated from those 
islands shown in Fig.  \ref{clamap}b  will increase their $\tilde{L}$ by 
$\pm\pi$ approximately, although the phase space structures have a period of $2\pi$;
(5) There exist interesting extensions of MKR. For instance, consider
the cases in which the kicking potential is reversed after every $N>2$ kicks  where
$N=6, 10, 14 \cdots$.  Then, for $\kappa=(2l_{2}+1)\pi$,
trajectories emanating from $(\tilde{L}= (2l_{1}+1)\pi, \theta=\pm
\pi/2)$ are also transporting, with their shift of angular
momentum after each kick alternating between
different constant values. However, 
we have found that the transporting islands associated with these generalized cases
are extremely small and very sensitive to the value of $\kappa$.

\begin{figure}[ht]
\begin{center}
\epsfig{file=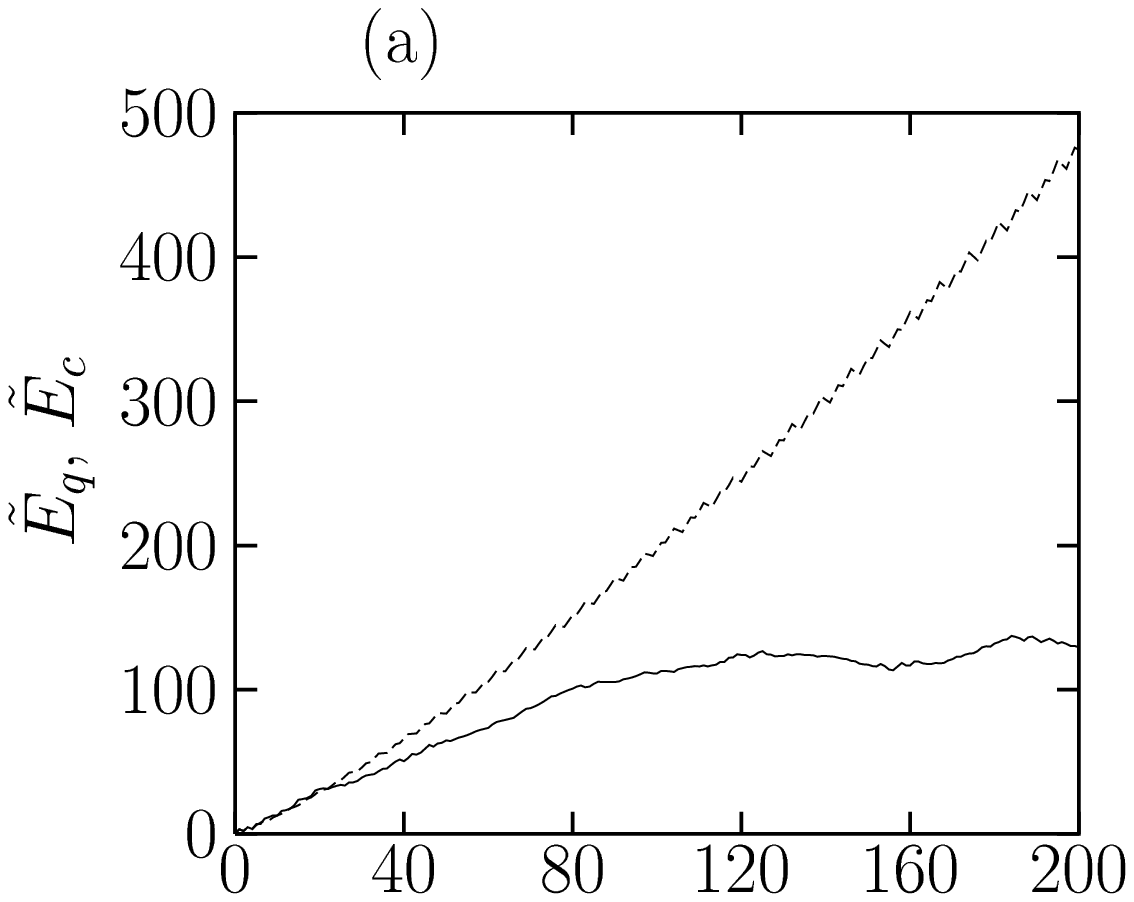,width=5.2cm}

\vspace{1.35cm}
\epsfig{file=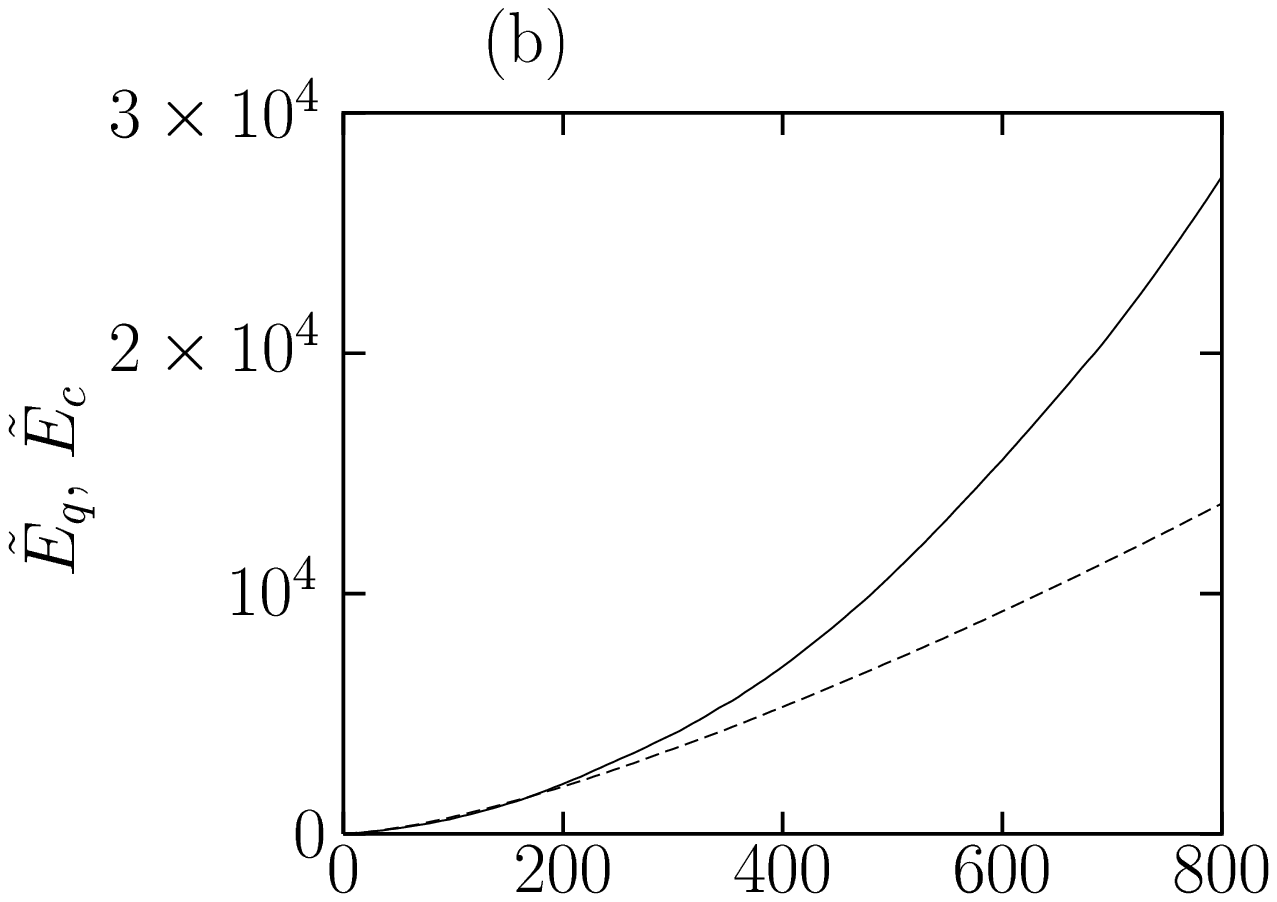,width=5.2cm}

\vspace{1.35cm}\epsfig{file=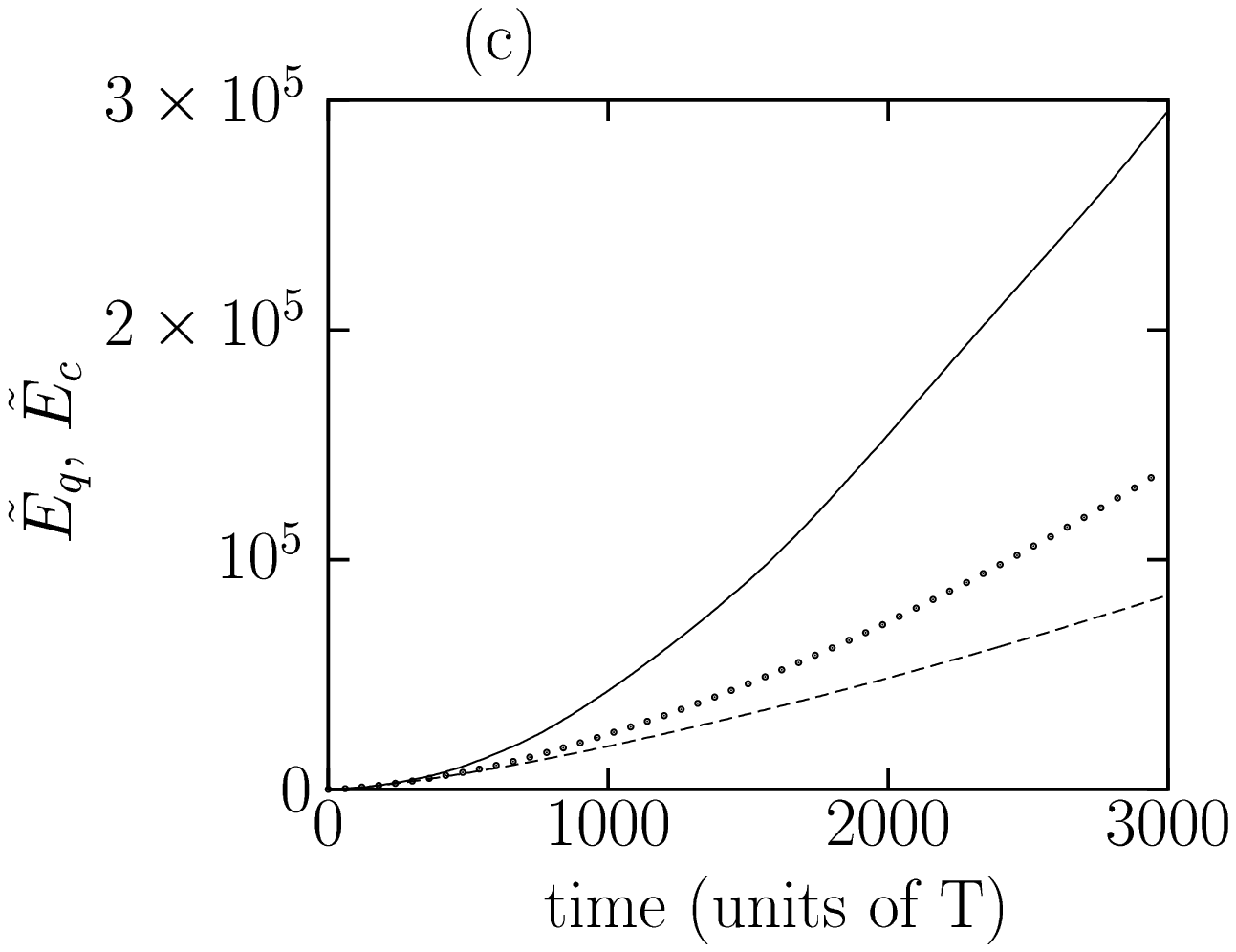,width=5.2cm}
\end{center}
\vspace{1cm}
\caption{The time dependence of the average dimensionless scaled
rotational energy, denoted $\tilde{E}_{q}$ and
$\tilde{E}_{c}$ for quantum and classical ensembles, respectively, for (a)
 KR, (b) MKR, and (c) MKR over even longer time scales, with $\kappa=3.5$.
  Solid lines denote quantum results for $\tau=0.1$,
   dashed lines denote classical results. The discrete points in (c)
    represent quantum results for $\tau=0.01$.
     Note that in (b) and (c) quantum results are well above the classical results.}
     \label{QC-compare}
     \end{figure}

Consider now quantum versus classical energy absorption behavior
in each of KR and MKR.  We choose $\kappa=3.5$ as an example;
the results described below
can be observed for a wide range of the parameter $\kappa$ 
(e.g., for $3.3<\kappa<5.0$).
We define the dimensionless scaled 
rotational energy averaged over the quantum or classical ensemble
as $\tilde{E}_{q}\equiv \langle \hat{L}^{2}\rangle\tau^{2}/2\hbar^{2}$ 
or $\tilde{E}_{c}\equiv \langle \tilde{L}^{2}\rangle/2$. 
The effective Planck constant $\tau$ is first chosen to be 0.1, a value far from
the semiclassical limit but
relatively small compared to the area of the transporting islands shown
in Fig. \ref{clamap}b. Further, 
this value of $\tau$ ensures that
the nongeneric behavior associated with quantum resonances (i.e., $\tau=2\pi l_{1}/l_{2}$) 
is avoided.  
The initial quantum state here is chosen to be $|0\rangle$, which
evidently does not overlap with the transporting islands.
The corresponding classical initial state is given by
$\tilde{L}=0$ with $\theta$ uniformly distributed in $[0, 2\pi]$. 
Figure \ref{QC-compare}a displays a quantum-classical comparison 
in the KR case.  Clearly, quantum effects in KR strongly
suppress classical energy absorption and
the QCC  break time  $t_{b}^{KR}$ in Fig. \ref{QC-compare}a  is 
$\sim 20T$. By contrast,
Fig. \ref{QC-compare}b shows that
both the quantum and classical MKR systems
display characteristics of anomalous diffusion:
an excellent $\log$-$\log$ linear fit
of the results in Fig. \ref{QC-compare}b after the MKR QCC break time $t_{b}^{MKR}$
[that is, we numerically fit the results with $\log(\tilde{E}_{q})=a\log(N)+b$ and
$\log(\tilde{E}_{c})=a'\log(N)+b'$, where $a$, $b$, $a'$,and $b'$ are parameters]
gives that $\tilde{E}_{q} \propto N^{1.85}$ (solid line)
and $\tilde{E}_{c}\propto N^{1.36}$ (dashed line),
where $N$ is the number of kicks. Hence, 
significantly, the quantum energy absorption is seen to be far greater than that associated with
the underlying classical
anomalous diffusion.  This rapid quantum growth is even more visible
in Fig. \ref{QC-compare}c, 
which shows that
the faster-than-classical excitation
process shown in Fig. \ref{QC-compare}b
persists for much longer time scales, as well as for
an effective Planck constant that is
ten times smaller (see the discrete points in Fig. \ref{QC-compare}c).
Note also that in the MKR case shown in Fig. \ref{QC-compare}b,
quantum 
dynamics agrees fairly well with the classical result for up to $t_{b}^{MKR}\sim 200T$,
much longer than $t_{b}^{KR}\sim 20T$ in Fig. \ref{QC-compare}a.
In addition,
$t_{b}^{MKR}$ is also  an order of magnitude larger than the
characteristic QCC time scale found
in the KR model in the presence of accelerator modes with the same
effective Planck constant \cite{grigolini}. This is understandable because 
the main transporting regular islands  
of MKR 
are larger than the accelerator modes
of KR and therefore
classical phase space structures can be ``seen" more clearly by the quantized MKR.

Also evident in Fig. \ref{QC-compare}c is that a smaller effective
Planck constant gives  
better  QCC.  
Thus, in contrast to what was suggested in Ref. \cite{grigolini},
the quantum anomalous diffusion here,   
approaches the underlying classical anomalous diffusion from above rather than from below
as the effective Planck constant goes to zero.
This leads to the conclusion  that under nonresonant conditions
quantization does not necessarily induce suppression of energy absorption.
Note that  one previous work 
\cite{lima} also reported computations (in the kicked-Harper model)
on faster-than-classical
diffusion under nonresonant conditions.
However,  in Ref. \cite{lima} the growth in the 
second moment of angular momentum is always quadratic 
and the issue of faster-than-classical
energy absorption cannot be addressed  since
the kinetic energy therein is a $\cos$-function of angular
momentum.
It should also be pointed out that, unlike 
gravity-induced quantum accelerator modes in KR \cite{schlunk},
quantum anomalous diffusion in MKR is shown to have
a well-defined classical limit
and does not require the value of $\tau$ to be 
near quantum resonance conditions.

It is interesting
to view MKR from a quantum control \cite{controlbooks} 
perspective.  
The time-evolving wave function can be expanded as a superposition of 
different $|m\rangle$ states, i.e., $\sum_{m} C_{m} \langle\theta|m\rangle$, where
$C_{m}$ are the expansion coefficients.  Since $\cos(\theta+\pi)=-\cos(\theta)$,
and $\sum C_{m} \langle\theta +\pi |m\rangle$=$\sum_{m} (-1)^{m}
C_{m} \langle\theta |m\rangle$, the effect of
reversing the kicking potential is equivalent to adding a
$\pi$ phase difference between all neighboring basis states.
As such, the potential reversal in
MKR can be regarded as periodically introducing quantum phases into KR 
and therefore as a significant extension of our previous coherent control
work \cite{gong01}. In addition, noting that
\begin{eqnarray}
\hat{F}^{MKR}(\tau,k)=\left[\exp(i\pi\frac{\partial^{2}}{\partial\theta^{2}})
\hat{F}^{KR}(\tau,k)
\hat{F}^{KR}(\tau,k)\right]^{2}, \end{eqnarray}
we have that MKR can be effectively realized
by using KR and introducing a time delay
$t_{d}=2\pi T/\tau$ of the kicking field after every two kicks.

\begin{figure}[ht]
\vspace{2cm}
\begin{center}
\epsfig{file=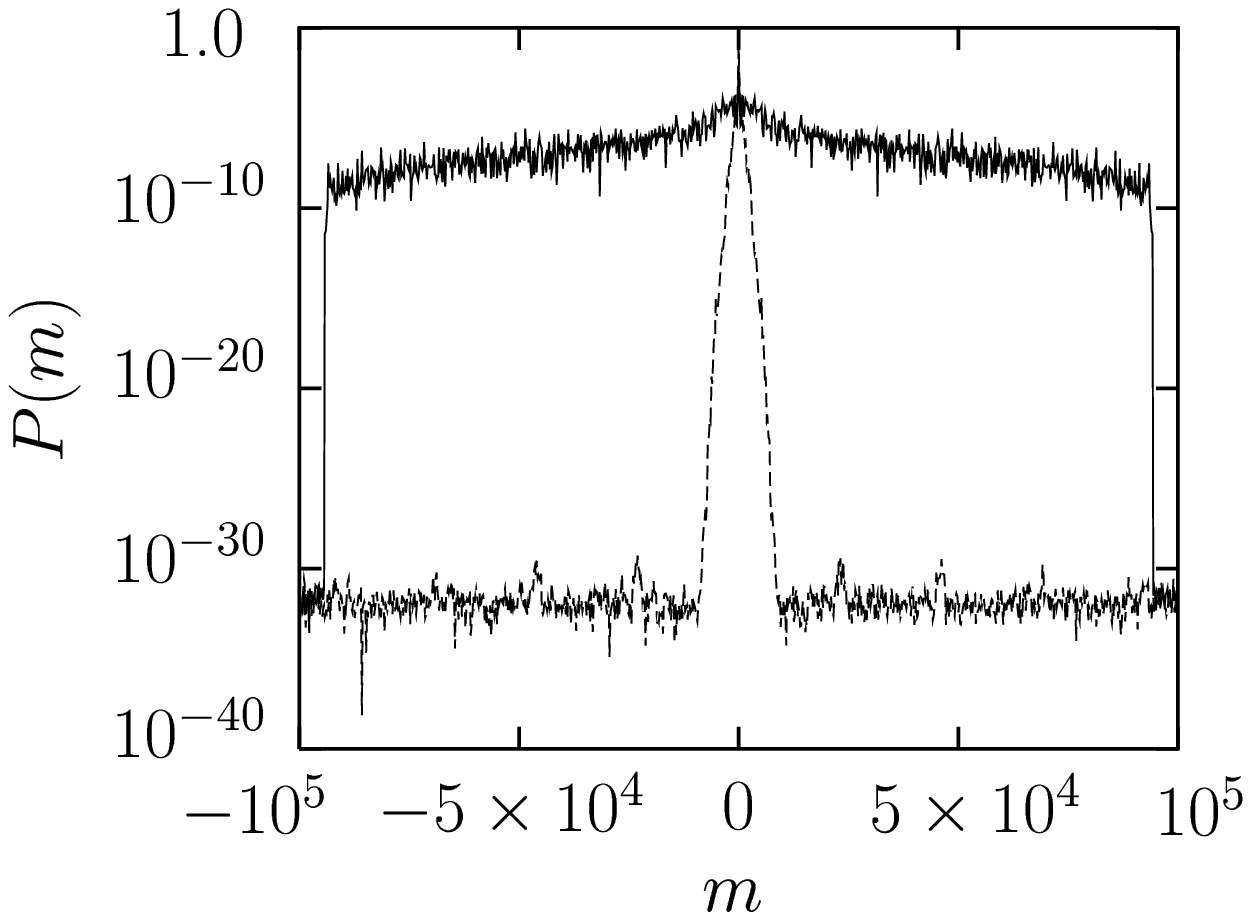,width=5.5cm}
\end{center}
\vspace{1cm}
\caption{A comparison between KR (dashed line) and MKR (solid line)
in terms of the probability $P(m)$
of finding the system in the state $|m\rangle$ at $t=3000T$. $\kappa=3.5$, $\tau=0.1$, and
the initial state is $|0\rangle$.
Fluctuations of $P(m)$ below the $10^{-30}$ level are due to numerical cutoff errors.}
\label{control}
\end{figure}

Figure \ref{control} displays a comparison between KR and MKR in terms
of the occupation probability $P(m)$  
of $|m\rangle$ after 3000 kicks, with $\kappa=3.5$,  $\tau=0.1$, and the initial state $|0\rangle$. 
It is seen that for many states
(e.g. for $90000>|m|>8000$), $P(m)$ of
MKR is greater than that of KR by a factor larger than $10^{20}$!
Further, we find that $\tilde{E}_{q}$ of MKR at $t=3000T$ is
three orders of magnitude larger than that of KR, whereas
classical dynamics only gives an energy-absorption difference of $\sim$ 
2.5 times. We stress (1) that 
this vast quantum control over energy absorption
is achieved by simply reversing the kicking potential, or,
alternatively, by introducing
a certain time delay after every two kicks, and (2) that the control mechanism
here is uniquely based upon quantum anomalous diffusion in MKR, and is thus far
more effective than
that in amplitude-modulated \cite{shepelyansky,casati} or phase-modulated
kicked-rotor systems \cite{gongpre03} in the absence of transporting islands.
We also note that if we regard MKR as controlled KR then we see
that we still have control for an effective Planck constant as large 
as $\tau=1.0$ \cite{gongpre03}.

\begin{figure}[ht]
\vspace{2cm}
\begin{center}
\epsfig{file=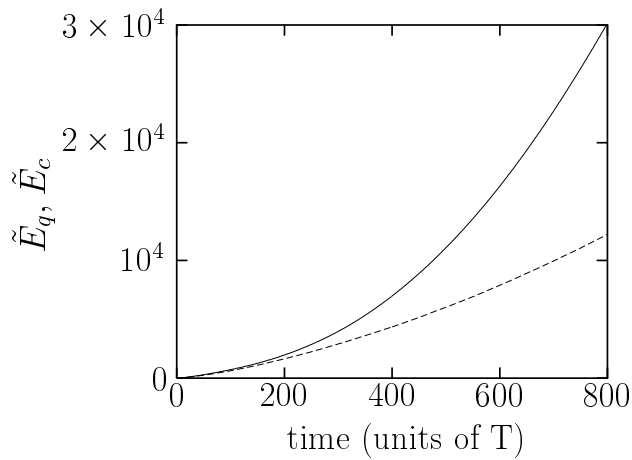,width=5.5cm}
\end{center}
\vspace{1cm}
\caption{As in Fig. \ref{QC-compare}b except that the initial condition, the
periodic boundary condition, the values of $\tau$ and $\kappa$ are all changed
(see the text for details).
The classical initial distribution function
is chosen to be identical with the Wigner function of the initial quantum state.
}
\label{all-changed}
\end{figure}

It remains to examine the possibility of observing faster-than-classical
quantum anomalous diffusion in atom optics experiments. 
In atom optics experiments on KR
the angular variable $\theta$ is replaced by the position variable of atoms
\cite{raizenetc}
so that there is no periodic boundary condition
as $|\psi(\theta=0)\rangle=|\psi(\theta=2\pi)\rangle$ in KR. Further,
in the experiments translational Gaussian states, which are different
from the initial state that we have examined thus far, are typically used as initial states.
Thus, of particular concern here is that the 
faster-than-classical quantum anomalous diffusion is found to be insensitive 
to the periodic boundary condition and to the initial conditions.
That is, in sharp contrast to quantum resonances in KR,
the faster-than-classical anomalous diffusion shown in Fig. \ref{QC-compare}b
and Fig. \ref{QC-compare}c
is found to be stable with respect to
small variations of all system parameters and initial conditions.
For example, Fig. \ref{all-changed}
shows similar faster-than-classical anomalous diffusion with 
the arbitrary choice of $\tau=2\pi/(60+\sigma)$, where $\sigma=(\sqrt{5}-1)/2$,
$k=33$, the initial state being 
given by a Gaussian $|\psi(\theta)\rangle=
\exp(-\theta^{2}/18)/\sqrt{3}\pi^{1/4}$, and the periodic boundary condition changed from
$|\psi(\theta=0)\rangle=|\psi(\theta=2\pi)\rangle$ to
$|\psi(\theta=0)\rangle=|\psi(\theta=512\pi)\rangle$.  This kind of 
result suggests that faster-than-classical quantum anomalous diffusion
should be observable in atom optics experiments, provided that 
the relatively small
effective Planck constant $\tau \sim 0.1 $ is achieved.

We qualitatively explain faster-than-classical anomalous diffusion in terms of
a quantum tunneling  mechanism \cite{zaslavsky02}.
At early times, both the quantum and classical systems that are 
not initially on the transporting regular islands tend to be
trapped by the fractal structures lying between the main transporting regular
islands and the chaotic sea, resulting
in excellent QCC in anomalous diffusion.
Later,  for $t \sim t_{b}^{MKR}$,
a non-negligible part of the time-evolving quantum state
has tunneled from
the chaotic sea to the large transporting islands, 
whereas each classical trajectory
can only sojourn in the neighborhood of the transporting islands for
a certain time.
Thus, if the mean sojourn time of classical trajectories
is relatively short compared to the characteristic time scale over which
the transporting-island component of
the time-evolving quantum state builds up and then tunnels back
to the chaotic sea,
the quantum tunneling effects
can strongly enhance
the transport in phase space, and therefore
quantum anomalous diffusion can be significantly faster than classical anomalous diffusion.
This also makes it clear that
quantum anomalous diffusion in MKR will slow down and dynamical
localization can take place when significant
tunneling in the reverse direction (i.e., 
from the transporting islands to the chaotic sea)  occurs. Indeed,
additional numerical results (not shown) for longer time 
scales (e.g., $12000$ kicks in the case of $\tau=0.1$) and for larger
effective Planck constants 
indicate that quantum anomalous diffusion in MKR will 
eventually be suppressed by
dynamical localization, with a larger $\tau$ giving earlier suppression.

In conclusion, we have found large transporting regular islands 
in the classical phase space of a modified
kicked-rotor system, and have shown that the associated
quantum anomalous diffusion can be significantly
faster than classical anomalous diffusion  over long time scales. 
The results are of interest to both areas of quantum chaos and
quantum control.

This work was supported by the U.S. Office of Naval Research and the
Natural Sciences and Engineering Research Council of Canada.
HJW was partially
supported by the Studien-stiftung des Deutschen Volkes and the Barth Fonds of ETH
Z\"{u}rich.

\end{document}